\documentclass{ifacconf}

\usepackage{natbib}        
\usepackage[hyphens]{url}
\usepackage{marginnote}
\usepackage[dvipsnames]{xcolor}
\usepackage{amsmath}
\usepackage{xfrac}
\usepackage{amssymb}
\usepackage{units}
\usepackage{graphicx}
\usepackage{booktabs}

\newcommand{\ddt}{\frac{\textnormal{d}}{\textnormal{d}t}}

\newif\ifmargincomments 
\margincommentstrue

\ifmargincomments

\else

\fi

\begin{document}
\begin{frontmatter}

\title{Electric Motor Design Optimization: \\A Convex Surrogate Modeling Approach\thanksref{footnoteinfo}} 

\thanks[footnoteinfo]{This publication is part of the project NEON (with project number 17628 of the research programme Crossover, which is (partly) financed by the Dutch Research Council (NWO)).}

\author[First]{Olaf Borsboom} 
\author[First]{Mauro Salazar}
\author[First]{Theo Hofman}

\address[First]{Eindhoven University of Technology, 
	5600 MB, Eindhoven, The~Netherlands (e-mail: \{o.j.t.borsboom, m.r.u.salazar, t.hofman\}@tue.nl).}

\begin{abstract}                
This paper instantiates a convex electric powertrain design optimization framework, bridging the gap between high-level powertrain sizing and low-level components design.
We focus on the electric motor and transmission of electric vehicles, using a scalable convex motor model based on surrogate modeling techniques.
Specifically, we first select relevant motor design variables and evaluate high-fidelity samples according to a predefined sampling plan. 
Second, using the sample data, we identify a convex model of the motor, which predicts its losses as a function of the operating point and the design parameters. 
We also identify models of the remaining components of the powertrain, namely a battery and a fixed-gear transmission.
Third, we frame the minimum-energy consumption design problem over a drive cycle as a second-order conic program that can be efficiently solved with optimality guarantees.
Finally, we showcase our framework in a case study for a compact family car and compute the optimal motor design and transmission ratio.
We validate the accuracy of our models with a high-fidelity simulation tool and calculate the drift in battery energy consumption.
We show that our model can capture the optimal operating line and the error in battery energy consumption is low.
Overall, our framework can provide electric motor design experts with useful starting points for further design optimization.
\end{abstract}


\end{frontmatter}

\section{Introduction}\label{sec:introduction}
\begin{figure}[t]
	\centering
	\includegraphics[width=\columnwidth]{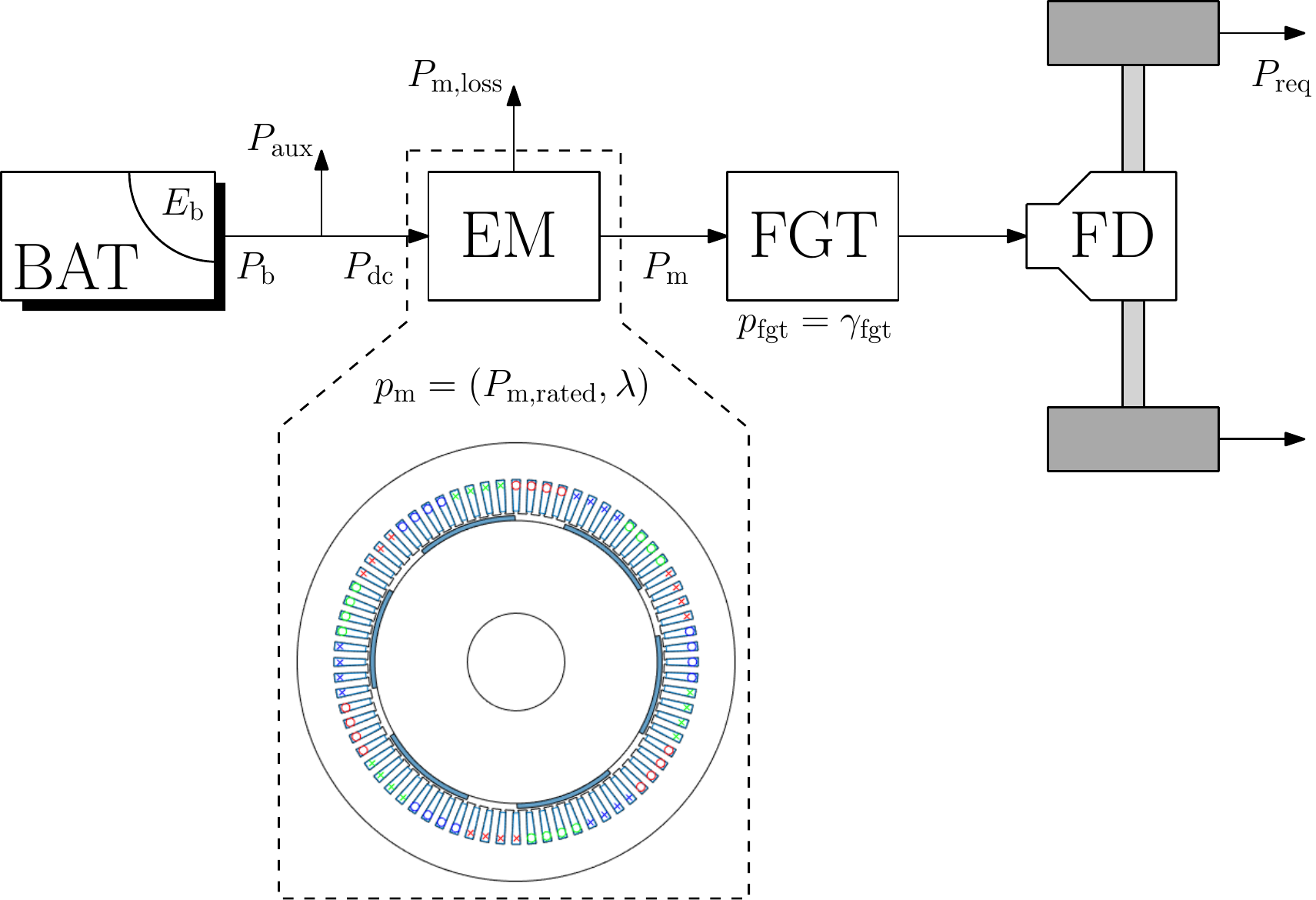}
	\caption{A schematic layout of the electric powertrain. It consists of a battery pack (BAT), an electric motor (EM), a fixed-gear transmission (FGT) and a final drive-differential unit (FD) that is connected to the wheels. 
	The arrows indicate the power flow between the components.	}
	\label{fig:powertrain}
\end{figure}

Electric vehicles are increasingly pervading the market, providing users with a zero-emission solution to personal mobility~\citep{IEA2021}.
However, to accelerate the widespread adoption of these vehicles, there is room for improvement in their affordability and range~\citep{PaoliGuel2022}.
Streamlining the design process of the electric \mbox{(e-)powertrain} is an important step towards this goal, which can be achieved by both reducing the time and cost of the technological development and converging towards better designs of the e-powertrain, accounting for the specific application.
This is a difficult task, since the e-powertrain is a complex system that consists of strongly coupled components, namely the battery, the electric motor (EM), the transmission, and the final drive-differential unit, as shown in Fig.~\ref{fig:powertrain}.
Moreover, there is typically a large disparity between the high-level vehicle requirements that powertrain designers might impose (in terms of performance, cost and energy efficiency) and the low-level component design questions that are raised in this context.
Specifically, optimizing the design of the EM has proven to be a challenge due to the high number of design variables and the multidisciplinary character of the problem~\citep{BramerdorferTapiaEtAl2018}.
To this day, in holistic powertrain design problems, the modeling is generally performed by using significant simplifications and assumptions to ensure the problem is computationally tractable. 
This is achieved, for instance, by linearly scaling the EM and its losses in the maximum torque and the mass, sacrificing accuracy~\citep{SilvasHofmanEtAl2016}.
However, implementing an optimization algorithm using accurate but computationally expensive models such as the finite element (FE) method, is not amenable to optimization.

This call for methods optimize the design of the e-powertrain with high accuracy, bridging the gap between high system-level sizing and low system-level design optimization of components.
We need to account for a scaling of the components with more relevant parameters, accuracy and wider scaling ranges, whilst maintaining computational tractability.

Against this backdrop, this paper presents a convex design optimization framework that leverages a scalable, convex EM optimization model based on surrogate modeling techniques.

\emph{Related literature:} 
This work relates to three main research streams. 
The first stream considers the design optimization of EMs on a low system level.
This issue is usually solved using FE-based approaches together with derivative-free optimization, but generally has a low-level objective, such as optimizing total harmonic distortion, torque ripple or power density~\citep{LeiZhuEtAl2017,BramerdorferTapiaEtAl2018,HanDangEtAl2017}. 
This does not connect well to the system-specific application (vehicle propulsion) and objective (powertrain energy consumption and cost).

The second steam considers the optimal sizing and control of (hybrid-)electric powertrains on a high system level. 
This problem is mainly addressed with derivative-free algorithms~\citep{EbbesenDoenitzEtAl2012,HegazyMierlo2010} or convex optimization~\citep{MurgovskiJohannessonEtAl2012,BorsboomFahdzyanaEtAl2021,SilvasHofmanEtAl2016}.
However, as mentioned earlier, these methods require simplified scalable models for complex components that are usually only valid for limited scaling ranges ($\pm$10-20\% w.r.t. the maximum torque).

The final stream aims to connect the previous two research streams and comprises the design optimization of powertrains with more detailed EM models. 
The methods are based on either parametric regression models of high-fidelity data in a relatively large EM training set~\citep{Zhao2017}, geometric scaling of a referent EM model~\citep{RamakrishnanStipeticEtAl2018}, or analytical design approaches~\citep{KruegerKeinprechtEtAl2022}.
However, all these methods lack global optimality guarantees and deal with high computational times.
What is more, relatively large training sets require much effort from EM design engineers, whereas analytical models are specified for a particular EM~\citep{HofmanSalazar2020}.

In conclusion, to the best of the authors’ knowledge, there are hardly any methods accounting for accurate EM design models in e-powertrain optimization in a computationally-efficient manner, giving accurate predictions over the whole design space, whilst limiting the efforts required by an EM design expert and still providing globally optimal solutions.

\emph{Statement of contributions:}
In order to address these challenges, this paper presents a convex optimization framework that optimizes the design of the EM and the transmission, based on surrogate modeling techniques.
Specifically, we first derive a scalable, convex EM model that predicts the losses as a function of the geometric dimensions and the rated power, trained with data from a pre-defined sampling plan.
Second, we leverage second-order conic programming to frame the minimum-energy consumption design problem, which minimizes the battery energy consumption of the e-powertrain over a drive cycle and computes the optimal EM design and transmission ratio.
Third, to showcase our framework, we solve the problem on the WLTP cycle using nonlinear numerical solvers, providing a solution guaranteed to be globally optimal.
Finally, we validate the accuracy of our solution with high-fidelity data.

\emph{Organization:} 
This paper is organized as follows:
Section~\ref{sec:methodology} presents the EM surrogate model and the encompassing optimization problem, which contains models and constraint functions for the vehicle and the remaining powertrain components.
We display our optimization framework in Section~\ref{sec:results}, after which we draw the conclusions in Section~\ref{sec:conclusions}, along with an outlook on future research.

\section{Methodology}\label{sec:methodology}
In this section, we construct the optimization problem by presenting convex constraints that describe the electric vehicle and its powertrain.
We present the objective and the vehicle and component models in Sections~\ref{subsec:objective}--\ref{subsec:battery} whereby we lay particular emphasis on the EM model, after which we summarize the problem in Section~\ref{subsec:ocp}, followed by a discussion on the assumptions in Section~\ref{subsec:discussion}.

The powertrain we consider in this paper, shown in Fig.~\ref{fig:powertrain}, contains a battery, an EM, and a fixed-gear transmission (FGT) that is connected to the wheels via a final drive and a differential. 
In this work, we are jointly optimizing the design of the EM and the FGT.

\subsection{Objective}\label{subsec:objective}
The objective in our optimization problem is to minimize the internal energy consumption of the battery over a drive cycle:
\begin{equation}\label{eq:objective}
		\!\min  \mathit{\Delta} E_\mathrm{b},
\end{equation}
where $\mathit{\Delta} E_\mathrm{b}$ is equal to the difference in battery state-of-energy (SOE), given by
\begin{equation}\label{eq:DeltaEb}
	\mathit{\Delta} E_\mathrm{b} = E_\mathrm{b}(0) - E_\mathrm{b}(T),
\end{equation}
where $E_\mathrm{b}(0)$ and $E_\mathrm{b}(T)$ are the SOE at the beginning and the end of the drive cycle, respectively.

\subsection{Longitudinal Vehicle Dynamics and Transmission}\label{subsec:vehicle}
In this section, we model the vehicle and the transmission.
As is common practice in powertrain sizing studies, we adopt the quasi-static modeling approach~(\cite{GuzzellaSciarretta2007}) in time domain.
To keep our derivations succinct, we will drop the time dependence $(t)$ whenever it is clear from the context.
The power requested at the wheels $P_\mathrm{req}$ is equal to
\begin{multline*}
	P_\mathrm{req} = v \cdot \left( \frac{1}{2} \cdot \rho_\mathrm{a} \cdot c_\mathrm{d} \cdot A_\mathrm{f} \cdot v^2 + \right. \\ 
	 \left. \vphantom{\frac{1}{2}} m_\mathrm{v}\cdot (g \cdot c_\mathrm{r} \cdot \cos(\alpha) + g \cdot \sin (\alpha) + a) \right),
\end{multline*}
where $v$, $a$ and $\alpha$ are the velocity, acceleration and road inclination given by the drive cycle, respectively, $\rho_\mathrm{a}$ is the density of air, $c_\mathrm{d}$ is the drag coefficient, $A_\mathrm{f}$ is the frontal area, $m_\mathrm{v}$ is the mass of the vehicle, $g$ is the Earth's gravitational constant, and $c_\mathrm{r}$ is the rolling resistance coefficient.
We assume that the design of the EM does not significantly influence the total mass of the vehicle $m_\mathrm{v}$, therefore we can compute $P_\mathrm{req}$ prior to the optimization.

We also assume that the FGT and the final drive have a constant efficiency. 
We only consider motor designs that can deliver the requested power, and we saturate the negative requested power with the maximum motor power $P_\mathrm{m,rated}$.
The mechanical motor power $P_\mathrm{m}$, which we can also pre-compute, is then equal to
\begin{align*}
	P_\mathrm{m} = 
	\begin{cases}
		\frac{1}{\eta_\mathrm{fgt} \cdot \eta_\mathrm{fd}} \cdot P_\mathrm{req}  & \text{if } P_\mathrm{req} \geq 0 \\
		\max(-P_\mathrm{m,rated}, \eta_\mathrm{fgt} \cdot \eta_\mathrm{fd} \cdot r_\mathrm{b} \cdot P_\mathrm{req})  & \text{if } P_\mathrm{req} < 0,
	\end{cases}
\end{align*}
where $\eta_\mathrm{fgt}$ and $\eta_\mathrm{fd}$ are the efficiencies of the FGT and final drive, respectively, and
$r_\mathrm{b}$ is the regenerative braking fraction.

We optimize the FGT ratio $\gamma_\mathrm{fgt}$, which is bounded by
\begin{equation}
	\gamma_\mathrm{fgt} \in [\gamma_\mathrm{fgt}^\mathrm{min}, \gamma_\mathrm{fgt}^\mathrm{max}],
\end{equation}
where ${(\cdot)^\mathrm{min}}$ and ${(\cdot)^\mathrm{max}}$ are the minimum and maximum values of the design variables.
The input speed of the transmission, which is identical to the output speed of the EM $\omega_\mathrm{m}$, is equal to 
\begin{equation}
	\omega_\mathrm{m} = \gamma_\mathrm{fgt} \cdot \gamma_\mathrm{fd} \cdot \frac{v}{r_\mathrm{w}},
\end{equation}
where $\gamma_\mathrm{fd}$ is the ratio of the final drive and $r_\mathrm{w}$ is the radius of the wheels.
We ensure the vehicle can reach the maximum velocity $v_\mathrm{max}$ by
\begin{equation}
	\gamma_\mathrm{fgt} \leq \omega_\mathrm{m,max} \cdot \frac{r_\mathrm{w}}{v_\mathrm{max}},
\end{equation}
where $\omega_\mathrm{m,max}$ is the maximum speed of the motor.
We also require the vehicle to be able to launch from standstill on a road inclination angle $\alpha$ by the following constraint:
\begin{equation}
	\gamma_\mathrm{fgt} \geq m_\mathrm{v} \cdot g \cdot r_\mathrm{w} \cdot \sin(\alpha_\mathrm{max})\cdot \frac{1}{\eta_\mathrm{fgt}\cdot \eta_\mathrm{fd}\cdot T_\mathrm{m,max}},
\end{equation}
where $T_\mathrm{m,max}$ is the maximum torque of the motor.

\subsection{Electric Motor}\label{subsec:em}
In this section, we derive a model of the EM.
As mentioned in Section~\ref{sec:introduction}, this work focuses on creating an accurate scalable model of the EM, whereby we draw inspiration from classical surrogate modeling techniques, whilst preserving convexity.
The goal of the EM model is to predict the motor losses $P_\mathrm{m,loss}$ as a function of the operating point and the design variables
\begin{equation*}
	P_\mathrm{m,loss} = f(P_\mathrm{m}, \omega_\mathrm{m}, p_\mathrm{m}),
\end{equation*}
where $p_\mathrm{m}$ is the set of EM design variables.

To construct the surrogate model, we require high-fidelity samples of EMs.
To this end, we use the open-source analytical tool MEAPA by~\cite{KaltEhrhardEtAl2020}, which is developed for the design and analysis of permanent magnet synchronous and asynchronous induction motors.
In this work, we focus on surface-mounted permanent magnet motors, with a fixed rated voltage $U$, rated speed $\omega_\mathrm{m,rated}$ and number of pole pairs $n_\mathrm{p}$.
The design variables we consider are the rated power $P_\mathrm{m,rated}$ and the relative length $\lambda$.
The relative length can be interpreted as the ratio between the length and the radius of the motor (measured at the stator's inner circumference between two poles), specifically given by
\begin{equation*}
	\lambda = \frac{l_\mathrm{s} \cdot 2\cdot n_\mathrm{p}}{\pi \cdot D_\mathrm{s,i}},
\end{equation*}
where $l_\mathrm{s}$ is the length and $D_\mathrm{s,i}$ is the inner diameter of the stator.

\subsubsection{Sampling Plan and Surrogate Model Formulation}
\begin{figure}[t]
	\centering
	\includegraphics[width=\columnwidth]{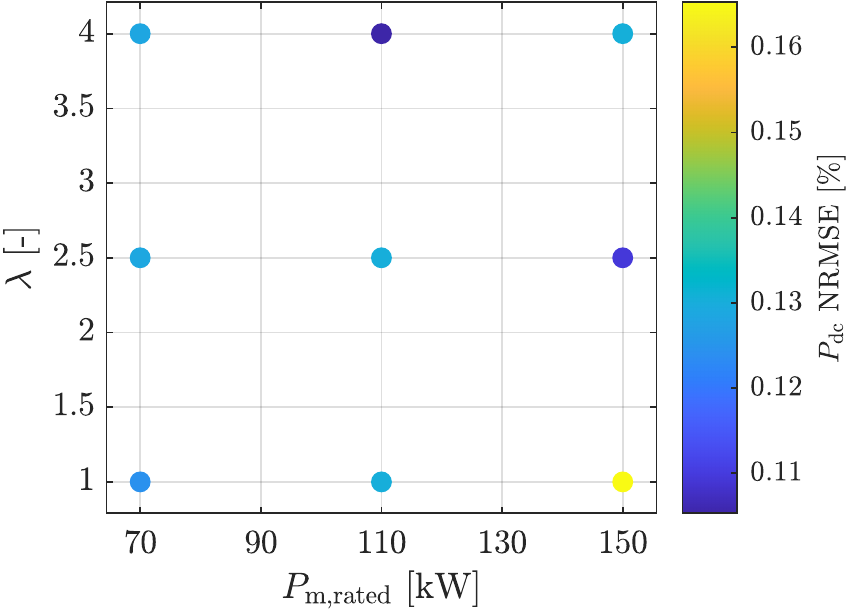}
	\caption{The EM sampling locations within the design space, evaluated with the MEAPA tool, following a 3-level 2-factor Full Factorial sampling plan. The NRMSE of $P_\mathrm{dc}$ for each sample is visible in the color bar, with the mean NRMSE for all samples equal to 0.13\%.}
	\label{fig:FFD_errors}
\end{figure}

To construct our scalable surrogate model, we perform high-fidelity evaluations on specific locations in the design space, according to a predefined sampling plan.
The design space is bounded by 
\begin{align}
	\lambda &\in [\lambda^\mathrm{min}, \lambda^\mathrm{max}] \\
	P_\mathrm{m,rated} &\in [P_\mathrm{m,rated}^\mathrm{min}, P_\mathrm{m,rated}^\mathrm{max}].
\end{align}
The specific sampling plan we select in this work, is a 3-level 2-factor Full Factorial sampling plan.
The locations of the samples in the design space are shown in Fig.~\ref{fig:FFD_errors}.

\begin{figure*}
	\centering
	\includegraphics[width=0.9\textwidth]{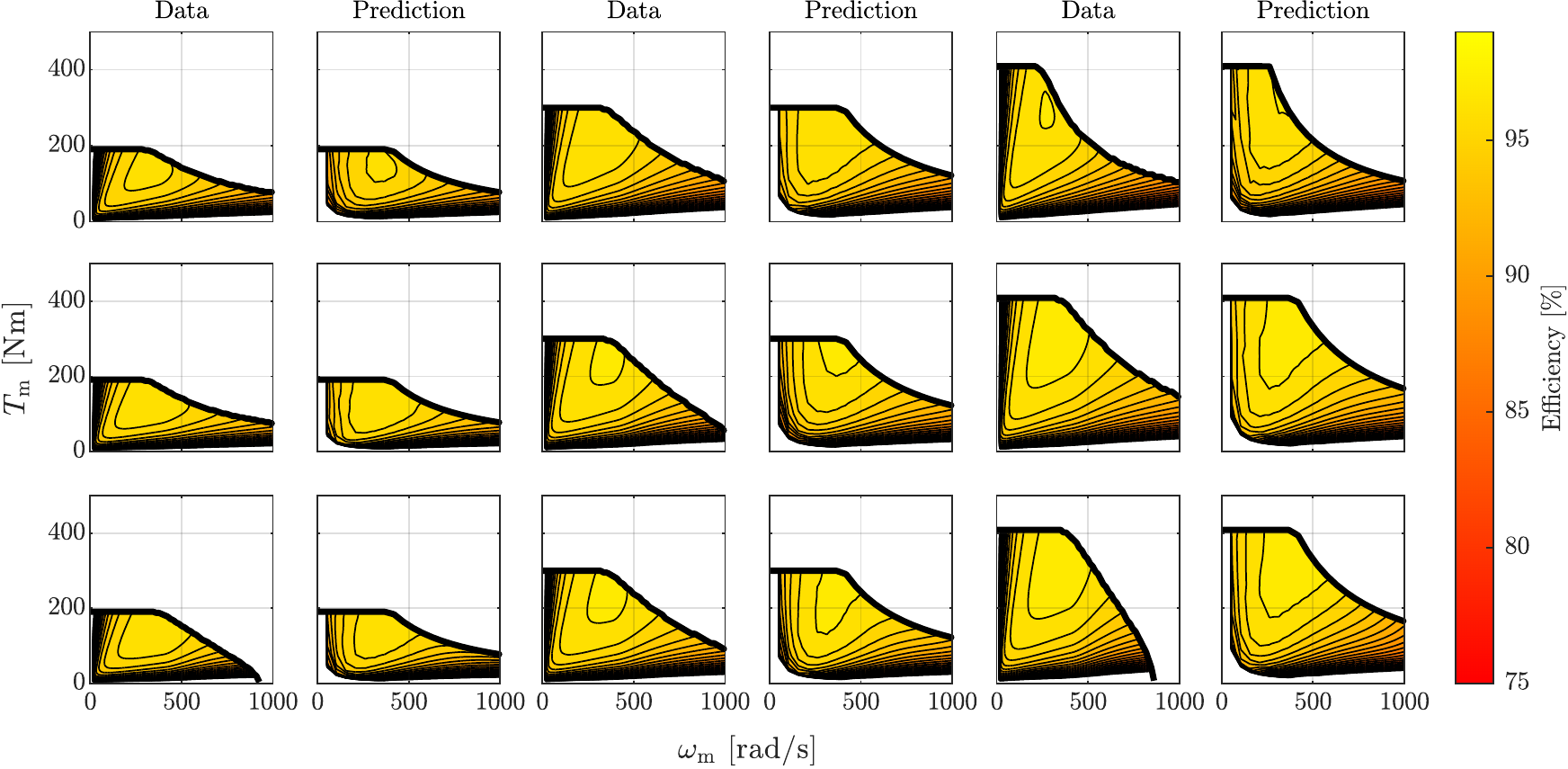}
	\caption{The efficiency maps of all sampled data points, along with the efficiencies predicted by the EM model.}
	\label{fig:FFD_eff_maps}
\end{figure*}

Using the high-fidelity data we acquired after evaluating the EMs in the sampling plan, we construct a convex, scalable surrogate model of the EM.
The EM model formulation is inspired by previous work, see~\cite{BorsboomFahdzyanaEtAl2021}, with the addition of design variables.
The electrical power $P_\mathrm{dc}$ is equal to
\begin{equation}
	P_\mathrm{dc} = P_\mathrm{m} + P_\mathrm{m,loss}.
\end{equation}
As mentioned before, we can pre-compute the required mechanical power exerted by the EM $P_\mathrm{m}$, hence we treat it as a given, exogenous parameter.
We train the model of the losses for $N$ different levels of $P_\mathrm{m} \in [0, P_\mathrm{m,rated}^\mathrm{min}]$, see~\cite{HurkSalazar2021,KorziliusBorsboomEtAl2021}.
We express the motor losses, after relaxing the constraint, as
\begin{equation}\label{eq:motorlosses}
	P_{\mathrm{m,loss},i} \geq x_\mathrm{m}^{\top} Q_{\mathrm{m},i} x_\mathrm{m}  \,
	\forall i \in [1,...,N],
\end{equation}%
where $Q_{\mathrm{m},i}$ is a matrix of fitting coefficients subject to identification, determined for each level of $P_\mathrm{m}$.
For power levels in between the fitted values, we linearly interpolate the fitting coefficients.
Vector $x_\mathrm{m}$, which contains constant values, main factors, first-order interactions, and quadratic terms, is equal to
\begin{multline*}
	x_\mathrm{m} = \left[
		\begin{matrix}
				1, & \omega_\mathrm{m}, & P_\mathrm{m,rated}, & \lambda, & \omega_\mathrm{m}\cdot P_\mathrm{m,rated}, & \omega_\mathrm{m}\cdot\lambda, \\
			\end{matrix} \right. 
		\\
		\left.
		\begin{matrix}
				 & P_\mathrm{m,rated} \cdot \lambda, & \omega_\mathrm{m}^2, & P_\mathrm{m,rated}^2, & \lambda^2
		\end{matrix}   
	\right].
\end{multline*}
To preserve convexity, we have relaxed the constraint in~\eqref{eq:motorlosses} and ensure that $Q_i$ are positive semi-definite matrices, see~\cite{Parrilo2004}.
Given the objective in~\eqref{eq:objective}, the constraint in~\eqref{eq:motorlosses} will always hold with equality.
After we determine values for the coefficients using semi-definite programming as in~\cite{BorsboomFahdzyanaEtAl2021}, we can assess the quality of the model by inserting the same design parameters as in the sampling plan for the operational envelope, up to $P_\mathrm{m}^\mathrm{min}$. 
The normalized root-mean-squared error (NRMSE) of predicting $P_\mathrm{dc}$ is visible for each sample in Fig.~\ref{fig:FFD_errors}, resulting in a mean NRMSE for all samples equal to 0.13\%.
The efficiency maps of the data points and the predictions are shown in Fig.~\ref{fig:FFD_eff_maps}.


We ensure the motor is powerful enough to complete the drive cycle with the following constraint:
\begin{equation}
	-P_\mathrm{m,rated} \leq P_\mathrm{m} \leq P_\mathrm{m,rated},
\end{equation}
We determine the maximum torque $T_\mathrm{m,max}$ from 
\begin{equation}\label{eq:Tmmax}
	T_\mathrm{m,max} \leq \frac{P_\mathrm{m,rated}}{\omega_\mathrm{m,rated}},
\end{equation}
which we implement into our framework, in both motoring and regenerating mode, by
\begin{align}
	T_\mathrm{m,max} \cdot \gamma_\mathrm{fgt} \cdot \eta_\mathrm{fgt}\cdot \eta_\mathrm{fd} &\geq \frac{P_\mathrm{req}\cdot r_\mathrm{w}}{v} \\
	T_\mathrm{m,max} \cdot \gamma_\mathrm{fgt} \cdot \frac{1}{\eta_\mathrm{fgt}\cdot \eta_\mathrm{fd}} &\geq \frac{P_\mathrm{req}\cdot r_\mathrm{w}}{v}.	
\end{align}
%
In constraint~\eqref{eq:Tmmax}, $T_\mathrm{m,max}$ can relax, but we allow it since it sets an upper bound for the torque of the motor.

\subsection{Battery}\label{subsec:battery}
In this section, we derive a model of the battery pack.
The output power of the battery $P_\mathrm{b}$ is equal to
\begin{equation}
	P_\mathrm{b} = P_\mathrm{dc} + P_\mathrm{aux},
\end{equation}
where $P_\mathrm{aux}$ is the auxiliary power.
Using the battery data from the quasi-static modeling toolbox in~\citep{GuzzellaSciarretta2007}, we model the internal battery power, after relaxation, as
\begin{equation}
	P_\mathrm{i} \geq b_0 + b_1\cdot P_\mathrm{b} + b_2\cdot P_\mathrm{b}^2,
\end{equation}
where $b_0$, $b_1$ and $b_2$ are parameters subject to identification, with a resulting NRMSE of 1.23\%.
The battery SOE changes with $P_\mathrm{i}$ as
\begin{equation}
	\ddt E_\mathrm{b} = -P_\mathrm{i}.
\end{equation}
We bound the battery SOE with the state-of-charge limits as
\begin{equation}
	E_\mathrm{b} \in [\zeta_\mathrm{b}^\mathrm{min}, \zeta_\mathrm{b}^\mathrm{max} ]\cdot E_\mathrm{b,max},
\end{equation}
where $E_\mathrm{b,max}$ is the predetermined total battery capacity---and thus not an optimization variable---and $\zeta_\mathrm{b}^\mathrm{min}$ and $\zeta_\mathrm{b}^\mathrm{max}$ are the minimum and maximum state-of-charge levels, respectively. 
We assume that the vehicle starts the drive cycle with a fully charged battery
\begin{equation}\label{eq:Eb0}
	E_\mathrm{b}(0) = \zeta_\mathrm{b}^\mathrm{max}  \cdot  E_\mathrm{b,max}.
\end{equation}

\subsection{Optimization Problem}\label{subsec:ocp}
In this section, we summarize the optimal design problem.
The state variable is $x = E_\mathrm{b}$.
The design variables are $p = \left(P_\mathrm{m,rated}, \lambda, \gamma_{\mathrm{fgt}}\right)$.

\begin{prob}[Nonlinear Convex Problem]\label{prob:main}
The minimum-energy design is the solution of
	\begin{equation*}
		\begin{aligned}
			&\!\min & &\mathit{\Delta} E_\mathrm{b} \\
			& \textnormal{s.t. } & &\eqref{eq:DeltaEb}-\eqref{eq:Eb0}.
		\end{aligned}
	\end{equation*}
\end{prob}
Although Problem~\ref{prob:main} is convex, it cannot be solved with standard convex solvers.
However, we can compute the solution, which is still guaranteed to be globally optimal, using nonlinear solvers.

\subsection{Discussion}\label{subsec:discussion}
A few comments are in order.
First, we assume that the mass of the EM does not significantly impact the total mass of the vehicle, enabling us to pre-compute $P_\mathrm{req}$.
In fact, the mass of the motor will change with $P_\mathrm{m,rated}$, yet its contribution to the total mass of the vehicle is relatively small and can be neglected~\citep{GrunditzThiringer2018}.
Second, in the training data set, some motors with a small value for $\lambda$ have to reduce the power at high rotational speeds in order to not exceed the limits of the circumferential rotor speed and the flux linkage.
However, since vehicles in this application hardly operate in this region of the envelope, it is considered of minor influence when optimizing the motor size in this stage of development.
Third, we assume a constant efficiency of the FGT, which is in line with common practice~\citep{VerbruggenSalazarEtAl2019}, and we assume that the cooling system can cope with the heating of the motor~\citep{KondaHofmanEtAl2022}.
Fourth, we have focused on two design variables for the EM, whilst the low-level design space for motors can be of a higher dimension.
Yet, our framework can be readily extended to account for more design parameters.
In the case that more design parameters are selected, the Full Factorial sampling plan---given its exponential characteristic in the number of parameters---can be replaced by Latin Hypercube, Central Composite or Box-Behnken sampling plans~\citep{GarudKarimiEtAl2017}.


\section{Results}\label{sec:results}
In this section, we present the numerical results obtained when we apply the optimization models presented in Section~\ref{sec:methodology} to the EM design of a compact car.
We optimize the design of the EM for the Worldwide Harmonized Light Vehicle Test Cycle (WLTC) Class 3.
Table~\ref{tab:vehparam} shows the vehicle parameters for which the optimal design is obtained.
The EM and FGT specifications are summarized in Table~\ref{tab:PTparam}.
Since our system dynamics ($E_\mathrm{b}$) are captured by an open integrator, we discretize the optimization problem with a sampling time of \unit[1]{s} using the forward Euler method.
In the case of closed-loop state dynamics (for instance in thermal modeling), other discretization methods could be necessary~\citep{LocatelloKondaEtAl2020}.
We parse the problem with CasADi~\citep{AnderssonGillisEtAl2019} and solve it with the nonlinear solver IPOPT~\citep{WachterBiegler2006}.
Because the problem is still convex, we preserve global optimality guarantees, since any KKT point found is a global minimum~\citep{BoydVandenberghe2004}.
Parsing the optimal design problem and solving it both take around \unit[7]{s}.
All computations are performed on an Intel Core i7-1065G7 CPU and \unit[16.0]{GB} of RAM.

\begin{table}[t!]
	\centering
	\caption{Vehicle Parameters}
	\label{tab:vehparam}
	\begin{tabular}{l l l l}\toprule
		\textbf{Parameter}   &   \textbf{Symbol}   &   \textbf{Value}   &   \textbf{Units}    \\ \midrule
		Wheel Radius         & $r_{\mathrm{w}}$    & 0.35               & [m]                 \\
		Air drag coefficient & $c_{\mathrm{d}}$    & 0.29               & [-]                 \\
		Frontal Area         & $A_{\mathrm{f}}$    & 2.38               & [m$^\mathrm{2}$]    \\
		Air density          & $\rho_{\mathrm{a}}$ & 1.2041             & [kg/m$^\mathrm{3}$] \\
		Rolling resistance coefficient & $c_{\mathrm{rr}}$ & 0.0174     & [-]                 \\
		Gravitational constant & $g$               & 9.81               & [m/s$^\mathrm{2}$]  \\
		Brake fraction       & $r_{\mathrm{b}}$    & 0.6                & [-]                 \\
		Final drive ratio    & $\gamma_{\mathrm{fd}}$ & 1               & [-]                 \\
		Vehicle mass         & $m_{\mathrm{v}}$   & 1,850                & [kg]                \\
		Auxiliary power 	& $P_\mathrm{aux}$ 		& 2 				& [kW] 				\\
		Maximum SoC          & $\zeta_{\mathrm{b}}^\mathrm{max}$ & 0.80          & [-]                \\
		Minimum SoC          & $\zeta_{\mathrm{b}}^\mathrm{min}$ & 0.20          & [-]                \\
	\end{tabular}
\end{table}

\begin{table}
	\centering
	\caption{Powertrain Parameters}
	\label{tab:PTparam}
	\begin{tabular}{l l l l}\toprule
		\textbf{Parameter}   &   \textbf{Symbol}   &   \textbf{Value}   &   \textbf{Units}    \\ \midrule
		\multicolumn{4}{c}{\textit{Electric Motor}}\\
		Voltage 			 & $U$                        & 700                 & [V] \\
		Rated speed 			& $\omega_\mathrm{m,rated}$      & 3,500 & [rpm] \\
		Maximum speed 		& $\omega_\mathrm{m,max}$ & 10,000 & [rpm] \\
		Number of pole pairs & $n_\mathrm{p}$   & 3             & [-]             \\
		Relative length bounds     & $\lambda^\mathrm{max}$     & 4          & [-]                \\
		& $\lambda^\mathrm{min}$ & 1          & [-]                \\
		                       
		Rated power bounds     & $P_\mathrm{m,rated}^\mathrm{max}$ & 150          & [kW]                \\
		& $P_\mathrm{m,rated}^\mathrm{min}$ & 70          & [kW]                \\

		\midrule
		\multicolumn{4}{c}{\textit{Fixed-gear Transmission}}\\
		Motor to Wheel Efficiency & $\eta_\mathrm{fgt} \cdot \eta_\mathrm{fd}$ & 0.96 & [-]     \\
		FGT ratio limits     & $\gamma_{\mathrm{fgt}}^\mathrm{max}$ & 10          & [-]                \\
		& $\gamma_{\mathrm{fgt}}^\mathrm{min}$ & 1          & [-]                \\
		Maximum velocity & $v_\mathrm{max}$ & 160 & [km/h] \\
		Maximum launch inclination & $\alpha_\mathrm{max}$ & 20 & [$^\circ$]	\\	
	\end{tabular}
\end{table}

\subsection{Numerical Results}
\begin{figure}[t]
	\centering
	\includegraphics[width=\columnwidth]{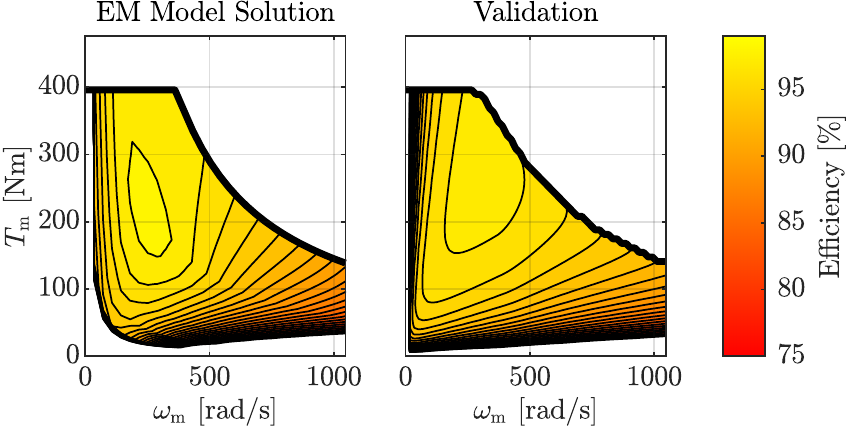}
	\caption{The efficiency map predicted by the EM model in the optimal solution ($P_\mathrm{m,rated} =$ \unit[145]{kW}, $\lambda =$ 3.49) is shown in the left subplot. On the right, the EM generated by the MEAPA tool for the same design as in the solution, serving as a design validation.}
	\label{fig:EMsolution}
\end{figure}

\begin{figure}[t]
	\centering
	\includegraphics[width=\columnwidth]{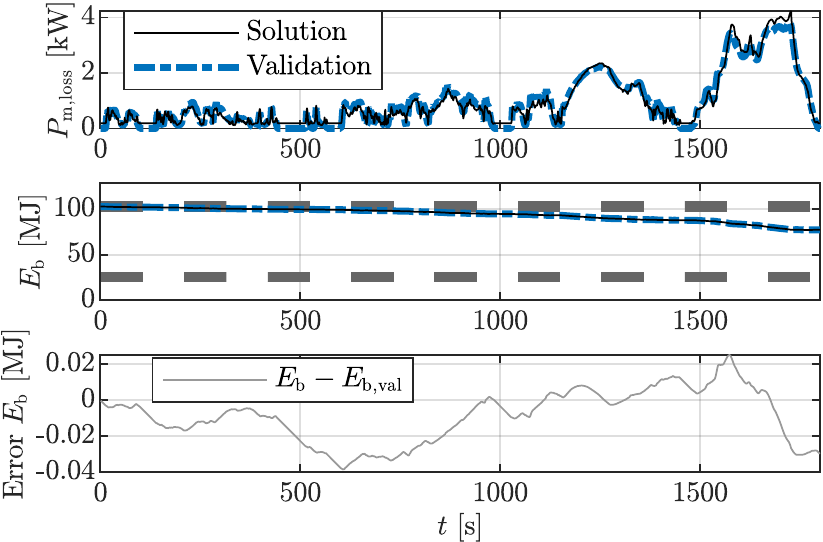}
	\caption{The trajectories of $P_\mathrm{m,loss}$ and $E_\mathrm{b}$, from both the optimal solution and validation. In the bottom subplot, the battery energy error in $E_\mathrm{b}$ is indicated.}
	\label{fig:solution_trajectories}
\end{figure}

After solving the optimal design problem for the parameters in Tables~\ref{tab:vehparam} and~\ref{tab:PTparam}, we arrive at an optimal design solution of $P_\mathrm{m,rated} =$ \unit[145]{kW}, $\lambda =$ 3.49, $\gamma_{\mathrm{fgt}} =$ 5.7. 
This is in line with current power ratings of electric vehicles with similar specification, such as the Volkswagen ID.3, which is rated at \unit[150]{kW}~\citep{Volkswagen2019}.
The efficiency map and limits predicted by the EM model of the solution design are shown in Fig.~\ref{fig:EMsolution}.
The predicted trajectories in terms of $P_\mathrm{m,loss}$ and the battery SOE $E_\mathrm{b}$ are shown in Fig.~\ref{fig:solution_trajectories}.


\subsection{Validation}
In order to validate our models, we feed the obtained design values in the MEAPA tool and carry out the analysis.
The resulting efficiency map is shown in the right subplot of Fig.~\ref{fig:EMsolution}.
Although the EM model prediction can more accurately capture the full efficiency map in the sampling points (Fig.~\ref{fig:FFD_eff_maps}), the EM prediction of the solution in the left subplot of Fig.~\ref{fig:EMsolution} can seize the optimal operating line relatively well.
To further quantify the modeling accuracy, we feed the obtained trajectories on the drive cycle from the solution through the nonlinear efficiency maps of the validation.
The resulting $P_\mathrm{m,loss}$ and the impact of it on $E_\mathrm{b}$ are shown in Fig.~\ref{fig:solution_trajectories}. 
The bottom subplot shows the error in battery energy consumption $E_\mathrm{b}$, which is quantified as the difference between the SOE trajectory from the solution ($E_\mathrm{b}$) and the validation ($E_\mathrm{b,val}$).
At the end of the drive cycle, the error is equal to \unit[30]{kJ}, which corresponds to 0.12\% w.r.t. the energy consumed.
This shows that the framework is capable of accurately predicting the losses of an optimally sized scalable motor.

\section{Conclusions}\label{sec:conclusions}
In this paper, we proposed to bridge the gap between high-level powertrain sizing and low-level EM design, and instantiated a convex optimization framework for the design of an electric motor (EM) and a fixed-gear transmission, which incorporates accurate scaling of the EM.
To this end, we took inspiration from surrogate modeling techniques and applied the ideas to predicting the losses of an EM as a function of its design, whilst preserving convexity.
After computing the energy-optimal design given a drive cycle, we compared the solution with the data obtained from the high-fidelity tool for the same design.
We observed that the predicted efficiency maps behave only slightly differently, the optimal operating line was captured well, with a relatively small error in energy consumption over a drive cycle of \unit[0.21]{\%}.
Therefore, this optimization model can aid EM design experts by providing them with a promising starting point, from which they can  further refine the low-level design of EMs for automotive applications.

This work opens the field to future research lines:
First of all, the accuracy of the model over the full design space---specifically, at the candidate optimum---could be improved by adding an iterative nature to the optimization procedure.
Namely, the use of infill points based on particular criteria can further explore the design space or refine the model in the solution~\citep{ForresterSobesterEtAl2008}.
Second, the model can be extended by including additional design variables.

\begin{ack}
We thank Dr. Ilse New for proofreading this paper.
\end{ack}

\bibliography{../../../Bibliography/main,../../../Bibliography/SML_papers}             

                                                   







\end{document}